\documentclass[journal=jacsat,manuscript=article]{achemso}
\setkeys{acs}{doi = true}

\usepackage[version=3]{mhchem} 
\usepackage{siunitx}

\usepackage{amssymb}
\usepackage{bm}
\usepackage{placeins}
\usepackage{glossaries}
\usepackage{booktabs}
\usepackage{graphicx}
\usepackage{etoolbox}
\usepackage{siunitx}
\robustify\itshape

\usepackage{color,soul}
\usepackage{xcolor}
\usepackage[colorlinks = true,
            linkcolor = blue,
            urlcolor  = blue,
            citecolor = blue,
            anchorcolor = blue]{hyperref}

\usepackage{makecell}

\newcommand{\doi}[1]{\href{http://dx.doi.org/#1}{\nolinkurl{#1}}}

\usepackage{etoolbox}

\makeatletter
\renewcommand*{\acs@author@fnsymbol@symbol}[1]{
    \ifcase #1 *\or
    1\or
    2\or
    3\or
    4\or
    5\or
    6\or
    7\or
    8\or
    9\or
    10
    \fi
}
        
\renewcommand*\acs@contact@details{
    {\sffamily *\,E-mail: \acs@email@list }%
    \acs@number@list
}           

\patchcmd{\acs@address@list@auxii}
{\acs@author@fnsymbol{\acs@affil@marker@cnt}}
{\textsuperscript{\acs@author@fnsymbol{\acs@affil@marker@cnt}}}
{}{}

\patchcmd{\acs@address@list@auxii}
{{\acs@author@fnsymbol{\acs@affil@marker@cnt}\@nameuse{@altaffil@\@roman\@tempcnta}\par}}
{{\textsuperscript{\acs@author@fnsymbol{\acs@affil@marker@cnt}}\@nameuse{@altaffil@\@roman\@tempcnta}\par}}
{}{}        

\makeatother

\author{Subhash V.S. Ganti}
\affiliation[1]{University of Bayreuth, Faculty of Engineering Science, Universitätsstraße 30, 95447 Bayreuth, Germany}
\alsoaffiliation[2]{University of Bayreuth, Bavarian Center for Battery Technology (BayBatt), Universitätsstraße 30, 95447 Bayreuth, Germany}
\author{Lukas Wölfel}
\affiliation[1]{University of Bayreuth, Faculty of Engineering Science, Universitätsstraße 30, 95447 Bayreuth, Germany}
\author{Christopher Kuenneth}
\affiliation[1]{University of Bayreuth, Faculty of Engineering Science, Universitätsstraße 30, 95447 Bayreuth, Germany}
\alsoaffiliation[2]{University of Bayreuth, Bavarian Center for Battery Technology (BayBatt), Universitätsstraße 30, 95447 Bayreuth, Germany}

\email{christopher.kuenneth@uni-bayreuth.de}

\title[]{AI-Driven Discovery of High Performance Polymer Electrodes for Next-Generation Batteries}

\begin{document}
\begin{abstract}

The use of transition group metals in electric batteries requires extensive usage of critical elements like lithium, cobalt and nickel, which poses significant environmental challenges. Replacing these metals with redox-active organic materials offers a promising alternative, thereby reducing the carbon footprint of batteries by one order of magnitude. However, this approach faces critical obstacles, including the limited availability of suitable redox-active organic materials and issues such as lower electronic conductivity, voltage, specific capacity, and long-term stability. To overcome the limitations for lower voltage and specific capacity, a machine learning (ML) driven battery informatics framework is developed and implemented. This framework utilizes an extensive battery dataset and advanced ML techniques to accelerate and enhance the identification, optimization, and design of redox-active organic materials. In this contribution, a data-fusion ML coupled meta learning model capable of predicting the battery properties, voltage and specific capacity, for various organic negative electrodes and charge carriers (positive electrode materials) combinations is presented. The ML models accelerate experimentation, facilitate the inverse design of battery materials, and identify suitable candidates from three extensive material libraries to advance sustainable energy-storage technologies.

\end{abstract}

\clearpage

\section{Introduction}

To reduce dependency on fossil fuels, electric batteries have become a prominent theme in modern scientific research \cite{li_30_2018,larcher_towards_2015}. However, several studies assert that the demand for electric vehicles (EV)  soars multifold, which is difficult to be met by transition metals only owing to the need of intensive mining and
depleting resource \cite{xu_future_2020,international_energy_agency_role_2021}. Currently, redox-active organic materials play an instrumental role in addressing the challenges posed by heavy reliance on fossil fuels and the sky-rocketing demand for transition-metal elements like Li, Co, and Ni in conventional batteries \cite{lu_prospects_2020,kim_redox-active_2024,esser_perspective_2021,bertaglia_eco-friendly_2024}. Redox-active organic materials exhibit diverse chemistries, structures, and applications for energy storage and mobility \cite{lu_prospects_2020}. They offer a wide range of characteristics like versatility, high-rate performance, and high theoretical capacity \cite{lee_redoxactive_2020}.

\begin{figure*}
 \hspace*{+0.75cm}
 \includegraphics{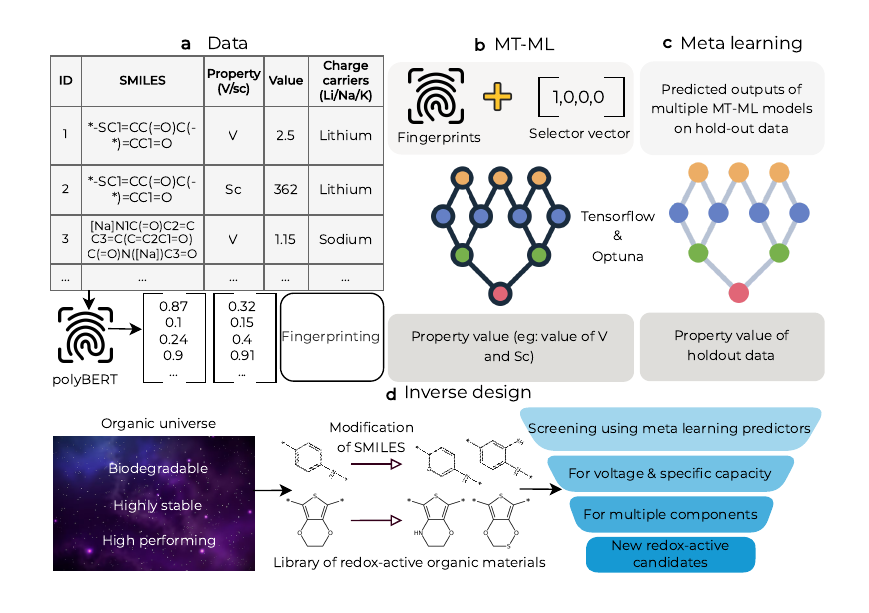}
 
  \caption{Our battery-informatics workflow. \textbf{a} Data collection pipeline: we collect SMILES (Simplified molecular input line entry system) strings of organic negative electrodes with respect to different charge carriers (positive electrode materials) \cite{weininger_smiles_1988}. This is followed by conversion of SMILES strings to numerical format. \textbf{b}  Architecture of our multi-task machine learning (MT-ML) predictors. MT-ML models are trained to predict multiple properties with respect to varying different battery components. Variation of charge carriers (positive electrode materials), property and organic material classes (polymers/molecules) is represented using selector-vector. The fingerprints are concatenated with selector vector and are used as inputs to MT-ML model. Lighter version of grey represents inputs of the model and darker version represents output. \textbf{c} Meta learners trained on holdout dataset by taking outputs of multiple MT-ML models as the inputs. The outputs of meta-learner models is the property value (voltage and specific capacity). \textbf{d} Finally, the inverse design approach is employed. We take some reference organic materials that exhibit either higher performance for batteries or higher stability or biodegradability. We iteratively add redox-active moieties or replace elements and bonds at different positions of the organic materials to create a library of millions of organic materials. We screen for potential candidates with higher voltage and specific capacity by using the proposed meta-learner model.}
  \label{fig:ml_pipline}
\end{figure*}

About two hundred redox-active organics are currently being used as electrodes in batteries \cite{bhosale_organic_2018,poizot_progress_2018,haupler_carbonyls_2015,schon_rise_2016,song_towards_2013,liang_organic_2012,wang_conjugated_2021}. However, they are bound by limitations such as dissolution in the electrolyte, poor electrical conductivity, or low volumetric density \cite{lee_recent_2018,shi_challenges_2022,yu_review_2024,kwon_versatile_2021}. To address these issues, we require novel organic materials with improved electrochemical performance \cite{li_electroactive_2022,yu_review_2024}. But the large chemical space of redox-active organic materials  \cite{reymond_chemical_2015,lu_perspectives_2023} makes it very time consuming and expensive \cite{reymond_chemical_2015,leelananda_computational_2016} for conventional approaches such as high throughput experimentation or combinatorial chemistry to identify possible candidates. ML methods provide a platform to navigate vast chemical spaces for designing materials with suitable properties \cite{zhang_transfer_2023,butler_setting_2024,schleder_dft_2019,shah_machine_2024,pai_machine_2025}. They are proven to be successful for material space explorations in the inorganic material space in the field of energy-storage applications (mainly batteries and super-capacitors) \cite{thakkar_advances_2024,chen_machine_2020,ahmad_machine_2018,shen_machine_2022}. In these ML models, compositional, elemental, and structural parameters are taken as ML inputs, while, for example, the discharge capacity is predicted \cite{thakkar_advances_2024,wang_machine-learning_2021}. 

For the unique and machine-readable representation of the chemical structures, Simplified Molecular Input Line Entry System (SMILES) strings are often used for encoding chemistries such as side chains, branches, rings, or chemical bonds \cite{weininger_smiles_1988}. Tools like polyBERT and Morgan fingerprints convert SMILES strings to vectors that serve as the numerical inputs to the ML models \cite{kuenneth_polybert_2023,bento_open_2020,rogers_extended-connectivity_2010}. Several previous investigations have utilized these fingerprinting approach for the prediction of diverse properties such as the HOMO-LUMO gap, atomization energy, redox potential, etc. \cite{ramakrishnan_quantum_2014,pinheiro_machine_2020,wu_moleculenet_2018,zhang_ring2vec_2024,schutt_quantum-chemical_2017,kuenneth_bioplastic_2022,kuenneth_copolymer_2021,shukla_polymer_2024}. Only limited research has focused on the prediction of experimental properties such as voltage for organic batteries \cite{carvalho_artificial_2022}. Moreover, while data scarcity entails a significant limitation for training ML models on experimental properties \cite{butler_machine_2018,chen_development_2024}, advanced learning approaches like multi-task, multi-fidelity, and transfer learning are contemporary solutions to address these challenges \cite{kuenneth_polymer_2021,hutchinson_overcoming_2017}. 

In this contribution, we employ a ML approach as illustrated in Figure \ref{fig:ml_pipline} to screen and identify high-voltage and high-specific capacity redox-active polymers for electrochemical applications. We utilize SMILES strings based notation to represent structures of organic materials (molecules and polymers) in our dataset. An important component of the workflow is the implementation of polyBERT as a fingerprinting tool, which transforms the SMILES strings into numerical representations suitable for ML model\cite{kuenneth_polybert_2023}. We develop a data-fusion MT-ML model followed by a meta learner to learn voltage and specific capacity for varying charge carriers (positive electrode materials). This multi step data-fusion methodology demonstrates improved generalizability and superior performance metrics, including better coefficient of determination (\textit{\SI{}{R\textsuperscript{2}}}) values and reduced RMSE (Root Mean-Square Error). Finally, we employ an inverse design methodology by creating three large libraries of polymeric materials. We propose new redox-active polymer candidates with maximum energy density, facilitating the efficient discovery and development of novel materials for battery applications.

\newpage
\section{Results and discussion}
\paragraph{Dataset}

\begin{table}
\begin{tabular}{{lrrrrrrrr}}
\toprule
{} & \multicolumn{2}{c}{Redox active} & \multicolumn{2}{c}{Non-redox active} & {} \\
{} & {V} & {Sc} & {V} & {Sc} & {} \\
{Charge carrier (+ve)} & {} & {} & {} & {} & {Total} \\
\midrule
\multicolumn{6}{c}{Polymer} \\
\midrule
Al & 2 & 4 & 0 & 0 & 6 \\
K & 1 & 1 & 0 & 0 & 2 \\
Li & 73 & 119 & 35 & 29 & 256 \\
Mg & 5 & 5 & 0 & 0 & 10 \\
Na & 10 & 7 & 29 & 35 & 81 \\
Zn & 2 & 3 & 0 & 0 & 5 \\
\midrule
\multicolumn{6}{c}{Molecule} \\
\midrule
Al & 1 & 0 & 0 & 0 & 1 \\
K & 2 & 3 & 0 & 0 & 5 \\
Li & 60 & 165 & 35 & 29 & 289 \\
Mg & 1 & 2 & 0 & 0 & 3 \\
Na & 19 & 27 & 29 & 35 & 110 \\
Zn & 3 & 0 & 0 & 0 & 3 \\
\midrule
Total & 179 & 336 & 128 & 128 & 771 \\
\bottomrule
\end{tabular}
\caption{Synopsis of the dataset for the battery property prediction models. The table shows the number of data points for different charge carriers (positive electrode materials) and polymer or molecule negative electrodes for voltage (V) and specific capacity (Sc).}
\label{dataset-table}
\end{table}

\begin{figure}[htbp]
    \centering
    \includegraphics{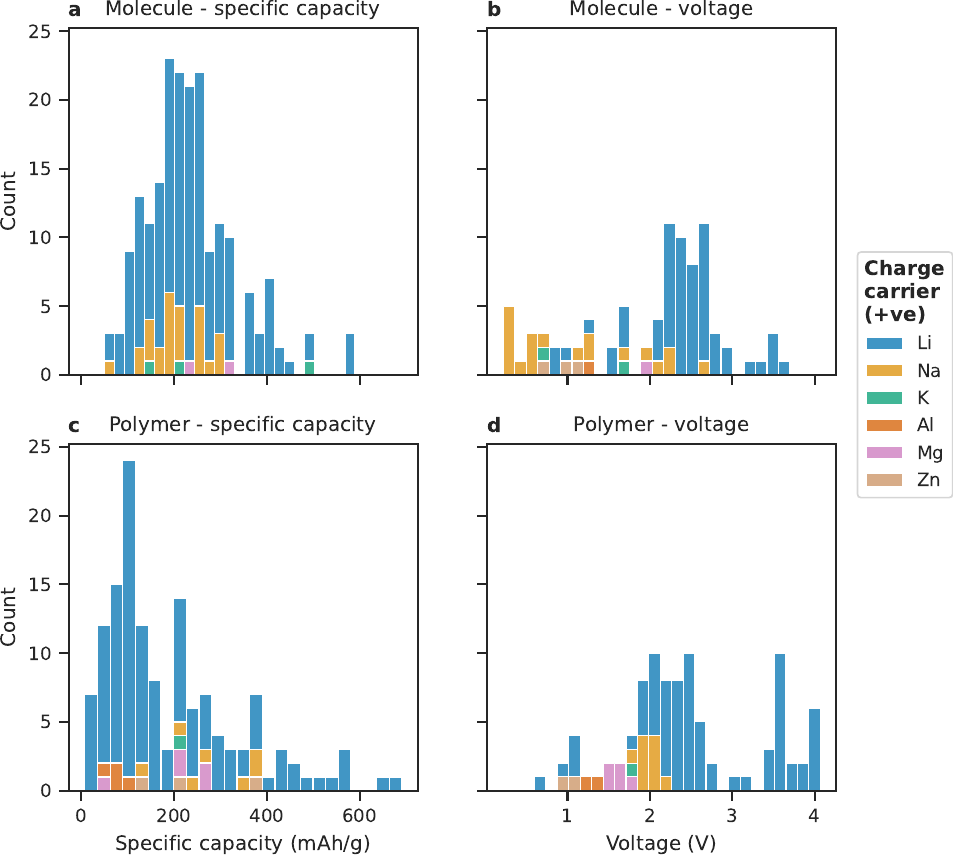}
    \caption{Stacked histogram plots for redox active molecules (panel \textbf{a} and \textbf{b}) and polymers (panel \textbf{c} and \textbf{d}) for specific capacity and voltage when used as negative electrodes with respect to different charge carriers (positive electrode materials) in batteries as indicated in the legend.}
    \label{fig:histogram}
\end{figure}

We utilize a dataset of a total of 771 data points for training our multi step data-fusion ML models as listed in Table \ref{dataset-table}. Our dataset contains both redox-active and non-redox active organic materials as negative electrodes with their electrochemical properties voltage in volts and specific capacity in \textit{\unit{\milli\ampere\hour\per\gram}}. For non-redox active materials, the dataset comprises of 128 data points for both properties, for a total of 256 data points. These data points are equally divided for both properties for lithium and sodium charge carriers (positive electrode materials) (85 points for each property and positive electrode) for molecules and polymers respectively. To ensure our models accurately reflect chemical reality, we include non-redox active materials in the training dataset. This teaches the models to differentiate between materials that can and cannot undergo redox reactions. On the other hand, our redox-active materials dataset is more extensive, containing 336 data points for specific capacity and 179 data points for voltage measurements, totaling 515 data points. It is worth noting that, 66 percent of data points (338 our of 506) in specific capacity are theoretically calculated. By incorporating these values, the ML models learn inherent correlations between theoretical and measured discharge capacity. In this process, the theoretical capacity is encoded using the selector vector (see Method section), allowing the model to effectively integrate and interpret the relationship between theoretical predictions and experimental measurements. This allows for the prediction of the costly and labor-intensive measured values directly from theoretical data, significantly enhancing the model's overall performance. The discharge specific capacity data points are collected with respect to varying C-rates and active material content for the first cycle. The distribution of both the properties for redox-active negative electrodes are shown in the figure \ref{fig:histogram}.

\paragraph{Model performance}
\begin{figure}[htbp]
    \centering
    \hspace*{-1cm}
    \includegraphics{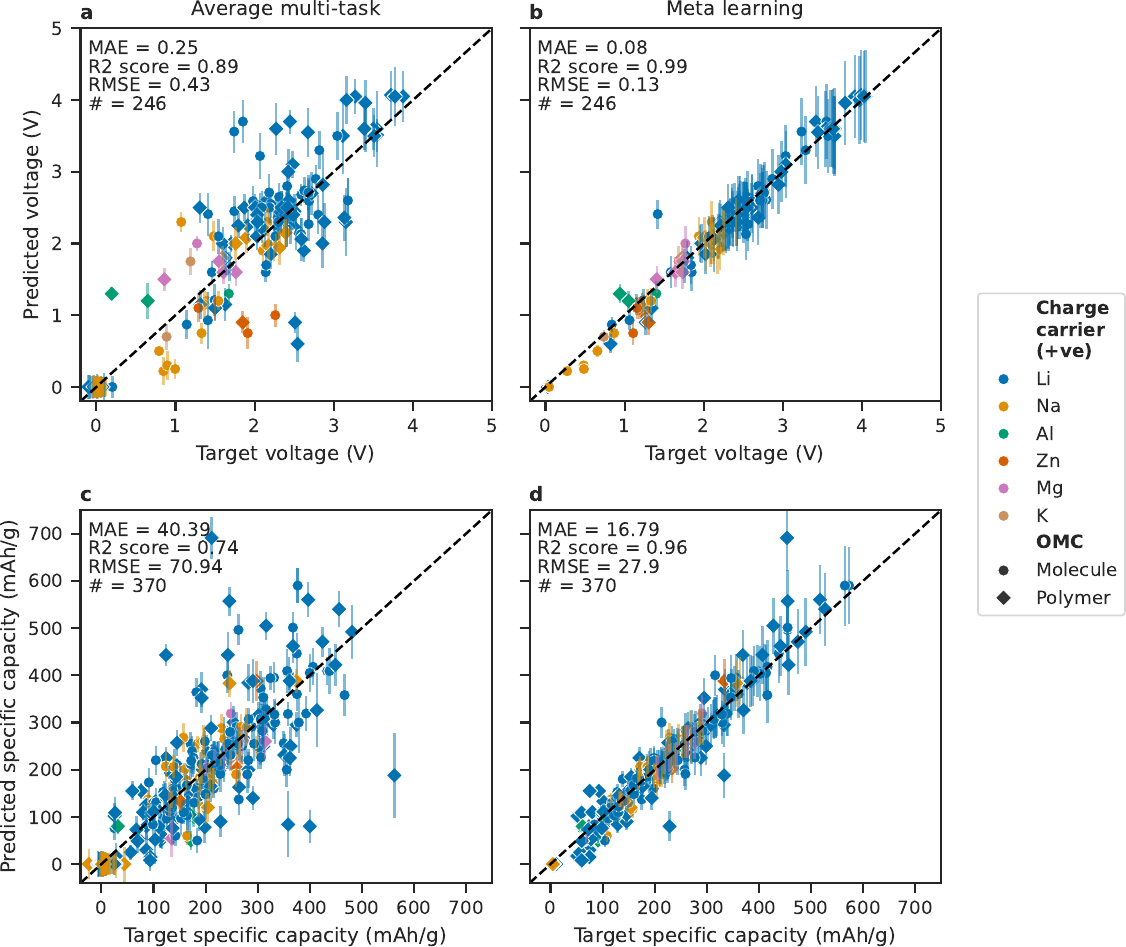}
    \caption{Parity plots for average of multi-task (five models) and meta-learner models for voltage and specific capacity with respect to varying charge carriers (positive electrode materials) and organic material classes (OMC) used. Plots \textbf{b} and \textbf{d} represent meta-learning whereas plots \textbf{a} and \textbf{c} represent average of testing sets of five-folds for MT-ML models. }
    \label{fig:enter-label}
\end{figure}
\DeclareSIUnit{\siga}{\ensuremath{\sigma}}
Utilizing our comprehensive dataset comprising of  771 data points, we train multi-task and meta-learning models of the prediction of voltage and specific capacity. Figure \ref{fig:enter-label} illustrates the comparative performance analysis of these models. The parity plots for MT-ML models, derived from the averaged results across all five-fold testing sets are shown in Figure \ref{fig:enter-label}\textbf{a} and \ref{fig:enter-label}\textbf{c}. The error bars indicate variance (\SI{2}{\siga}) of the predictions across the five-folds.  MT-ML models demonstrate their efficiency in accurately predicting multiple electrochemical properties.  A single MT-ML model is trained on multiple properties simultaneously, enhancing the predictive efficiency, addressing data scarcity and eliminating the necessity of training individual model for each property, as shown in Reference \cite{kuenneth_polymer_2021}. The MT-ML models achieves RMSE values of \SI{0.43}{\V} and \SI{70.9}{\milli\ampere\hour\per\gram} for voltage and specific capacity,respectively. The voltage predictions of lithium charge carrier and polymeric negative electrodes are closer to the parity line in Figure \ref{fig:enter-label}\textbf{a} when compared to their molecular counterparts. On the other hand, prediction of specific capacity perform better for molecular negative electrodes and lithium charge carrier in Figure \ref{fig:enter-label}\textbf{c}. For charge carriers containing lesser data-points, the predictions are spread on the both sides of the parity line. Our multi-task models exhibit robust performance across diverse electrode materials, successfully predicting properties for various charge carriers (positive electrode materials) and organic material classes (OMC).

Subsequently, we implement and train meta-learning techniques to enhance and bundle the predictive capabilities of our MT-ML framework  for the deployment of our ML models. The resulting parity plots in Figure \ref{fig:enter-label} demonstrate the exceptional generalization capacity of our meta-learner models, achieving remarkably high \SI{}{R\textsuperscript{2}} values of 0.99 and 0.95 for voltage and specific capacity, respectively.  A comparative analysis between multi-task (panel \textbf{a} and \textbf{c}) and meta-learner (panel \textbf{b} and \textbf{d}) in Figure \ref{fig:enter-label} results reveals two significant improvements: (i) enhanced alignment of predictions with the parity line and (ii) substantial improvement of predictions for charge carriers (positive electrode materials) with less data-points. Notably, our models has these exceptional performance metrics irrespective of the inherent challenges of working with a relatively small dataset and multiple fidelity levels (incorporating both theoretical and experimentally measured capacity values across varying C-rates and active material loadings as encoded in the fingerprints). The strong improvement in predictive accuracy from multi-task to meta-learner models showcases the efficiency of our multi step data-fusion approach in addressing the complexities in electrochemical property predictions.

\paragraph{Inverse design}

\begin{figure}
    \hspace*{-0.7cm}
    \includegraphics{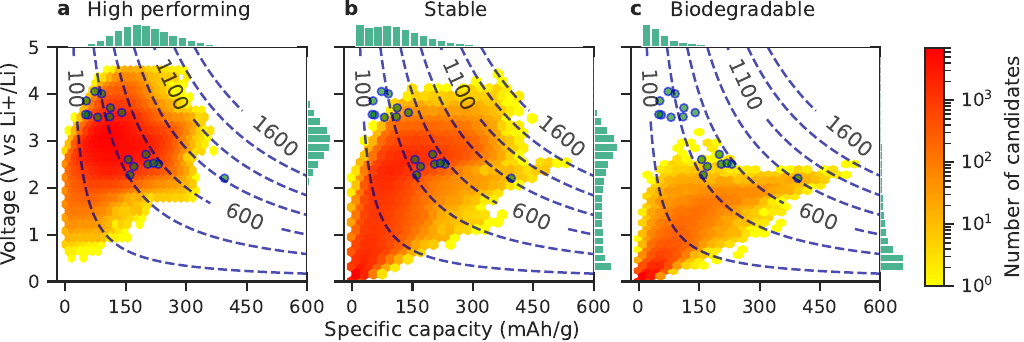}
    \caption{Specific capacity vs voltage distribution for three libraries totaling to 1.8 million candidates. The voltage values represent average working potential versus lithium, while specific capacity values correspond to discharge capacity at 1C for the first cycle (50 percent active material loading). The overlaid dashed lines represent energy density (\SI{}{\watt\hour\per\kilo\gram}).The blue colored points represent experimental data-points versus lithium at 1C used for training our property predictor models.}
    \label{fig:inverse_design_plot}
\end{figure}

In our inverse design methodology, we establish a search space from 11 selected reference organic polymers as shown in the Supplementary Figure 2. Among those 11, two polymers are chosen for their high voltage, six for their structural stability and the remaining two based on their biodegradability. As illustrated in Figure \ref{fig:inverse_design_plot}, the polymer candidates are categorized into three distinct libraries based on their reference: high-voltage polymeric negative electrodes (Fig. \ref{fig:inverse_design_plot}a), stable plastics (Fig. \ref{fig:inverse_design_plot}b), and biodegradable polymers (Fig. \ref{fig:inverse_design_plot}c). These candidate libraries serve as search space for potential replacements to existing organic electrodes that exhibit high voltage, specific capacity, or energy density. Through systematic structural modifications like strategic addition and substitution of redox-active moieties (see Supplementary Figure 3) at both main-chain and side-chain positions, the 11 reference polymers are expanded into an extensive library of approximately 1.8 million candidates. The electrochemical properties of these candidates were predicted using our two-step data-fusion models, with the resulting property distributions as depicted in Figure \ref{fig:inverse_design_plot}. The observed property distributions exhibit systematic correlations that reflect fundamental chemical principles through the direct modification of molecular electronic structure, which occurs via incorporation of electron-withdrawing or electron-donating redox-active groups in organic materials \cite{bitenc_organic_2024}. Our screening process identifies several promising candidates with operating voltages between \qtyrange{4.1}{4.5}{\volt} (see Supplementary Figure 5), exceeding the maximum voltage of \SI{4.07}{\volt} present in our training dataset.
Analysis of the distribution patterns reveals varying characteristics of the three libraries. The high-voltage organic negative electrode library (Fig. \ref{fig:inverse_design_plot}a) shows a notable concentration of candidates in the high-voltage regime with moderate specific capacity. Conversely, the biodegradable polymer library (Fig. \ref{fig:inverse_design_plot}c) exhibits an opposite trend (see Supplementary Figure 6). All-organic batteries manufactured from biodegradable redox-active polymers present a viable solution for disposable electronic devices, offering enhanced safety in post-usage disposal \cite{lee_biodegradable_2021}. The stable plastics library (Fig. \ref{fig:inverse_design_plot}b) achieves the highest energy density (see Supplementary Figure 4), approaching \SI{1600}{\watt\hour\per\kilo\gram}, indicating that the incorporation of redox-active moieties into stable organic backbones improves multiple battery properties \cite{liang_organic_2012}. On the other hand, our model also identifies materials with insulating properties, characterized by near-zero values across both properties, highlighting its unbiased prediction capabilities. This comprehensive performance spectrum, ranging from high-performing candidates to insulators, demonstrates the model's robust predictive capacity across the different classes of materials. Each of the three libraries—stable plastics, biodegradable materials, and high-performing polymers were screened for suitable candidates in an average of 1.5 minutes (90 seconds) per library. 

\section{Conclusion}

Our investigation demonstrates the effectiveness of our data fusion ML model for predicting electrochemical properties across multiple domains and fidelity levels (active material loading, C-rate, property (voltage or specific capacity), organic material classification (molecule or polymer), and the charge carrier in positive electrode (Li, Na, K, Zn, Mg and Al). The MT-ML framework is instrumental in addressing the critical challenge of data scarcity while enhancing the prediction accuracy simultaneously. Building upon this foundation, our meta-learning approach significantly improves model performance and generalization capabilities across various property domains and material classifications. Our framework shows significant computational efficiency, enabling ultra-fast prediction of electrochemical properties for large-scale material screening. This capability proves instrumental in the discovery of next-generation battery materials. By leveraging an inverse-design approach, our framework enables the rapid generation of candidate material libraries.  Furthermore, it suggests both promising novel candidates and effective replacements for existing materials, significantly accelerating the screening process and reducing the time and resources required for experimental validation.

Our research aims to enhance the accuracy and efficiency of organic-battery informatics approach by integrating the influence of electrolytes and separators. Additionally, we are developing novel fingerprinting approach for inorganic materials in batteries which has the potential to significantly increase the number of available data points, leveraging the extensive databases available for batteries in inorganic space. Our research has broader implications beyond theoretical advancement, facilitating the development of cost-effective and environmentally sustainable energy storage solutions. By paving the way for the discovery of high-performing organic battery materials, this work represents a significant step towards our goal of identification of all-organic batteries using AI. These advancements could lead to batteries with improved capacity, longer lifespans, and reduced environmental impact, addressing critical challenges in the field of energy storage.

\section{Methods}
\paragraph{Data Preparation}
Our dataset comprises of 771 data points curated from peer-reviewed research publications, prominently focusing on review articles \cite{kim_organic_2022,judez_energy_2019,liang_organic_2012,lu_prospects_2020}. We collected and curated the data from the review articles for different negative and charge carriers (positive electrode materials), voltages and specific capacities. The UMAP (Uniform Manifold Approximation and Projection) plot of our dataset visualizing distribution of different organic material classes (polymers and molecules) of the dataset in a two dimensional embedding space is shown in Supplementary Figure 1 \cite{mcinnes_umap_2018}. Prior to partitioning the dataset into training and testing subsets, we applied min-max scaling to normalize each target property (voltage and specific capacity) independently, linearly transforming the output values to a range based on their respective minimum and maximum values within the dataset.

\paragraph{Fingerprinting}
Fingerprinting involves conversion of chemical information represented by SMILES to machine-readable numerical format in form of vectors. For fingerprinting, we utilize the tool polyBERT \cite{kuenneth_polybert_2023}, resulting in 600-dimensional vectors. Conversion to fingerprinting is followed by concatenation with selector vectors. Selector vector is an indicator of distinct domains in the data. They are five distinct numerical values, which encode key experimental and material parameters: active material content, C-rate, property type, organic material class (molecule, polymer, or ladder polymer), and the charge carrier (e.g., Li, Na, K). An example of a selector vector used in the model is \bm{$[0.5, 1, 1, 1, 1]$} indicating that the organic material is a polymer with 50\% active material content, operating at 1C. Additionally, the property being predicted is voltage w.r.t a sodium-positive electrode. Since, while calculating theoretical capacity, C-rate cannot be mentioned, we represented it using the number -1. When fingerprints are concatenated with the selector vectors, the input to our ML models yields a 605-dimensional numerical representations. This approach enables the representation of multiple fidelity levels within a single computational framework, thereby reducing generalization error \cite{kuenneth_polymer_2021}.  

\paragraph{Model Architecture and Training Methodology}
Our two-step data-fusion ML approach encompasses both MT-ML and meta-learning components. The dataset was randomly split, allocating 80\% for MT-ML model development and reserving 20\% for meta-learner training. The MT-ML training dataset  is subdivided into 5-folds for cross validation, and 5 independent models are trained for each fold. The neural network-based MT-ML optimization is performed using the TensorFlow framework, supported by Optuna \cite{tensorflow2015-whitepaper,akiba_optuna_2019}, enabling systematic and full hyper-parameter tuning across neural network architecture parameters (i.e., neuron count, number of layers, activation function, and dropout rate) and training parameters (i.e., initial learning rate and early stopping criteria). The hyper-parameter optimization for each fold is performed independently.
The meta-learning step consists of a deployment-ready ensemble framework integrating insights deduced from all cross-validated models. The meta learning follows a two-step methodology: initially, predictions are generated for the held-out 20\% dataset using the 5-fold cross-validated models, followed by the training of a neural network that uses these predictions as inputs to learn final property values. The model's generalization capability is tested against the original 80\% training dataset. Comparison of cross-validated and meta learner models are shown in Supplementary Table 1 and Table 2. Uncertainty quantification is implemented for both multi-task and meta-learning predictions using Monte Carlo dropout methodology, providing confidence intervals at the 95\% level \cite{gal_dropout_2015}. This meta-learning approach ensures comprehensive model generalization across the entire dataset while maintaining prediction reliability.

\DeclareSIUnit{\COtwoeq}{CO_2eq}

\section{CO$_2$ Emission and Timing}
Experiments are conducted using a computing cluster at the University of Bayreuth, with the carbon efficiency of \SI{0.344}{\kilo\gram\COtwoeq\kilo\watt\per\hour}. A total of \SI{10}{\hour} of computations were carried out on four A-100-40GB GPU's with a Thermal Design Power (TDP) of \SI{250}{\watt}. The training of MT-ML and meta learner models producing emissions around \SI{1.4}{\kilo\gram\COtwoeq}. In the inference mode, the computation per polymer is around two seconds combining fingerprinting, concatenation of selector vector and prediction from the trained model in total on one GPU, emitting around \SI{57}{\gram\COtwoeq}.
\section{Data Availability}
The data that support the findings of this study are available from the authors upon reasonable request.
\section{Code Availability}
The code used for training the MT-ML and meta learner models are available for academic use at at \url{https://github.com/kuennethgroup/organic_battery_predictor} and Zenodo \cite{subhash_v_s_ganti_2025_14886803}.

\bibliography{references}

\section{Acknowledgements}
The authors sincerely thank the graduate school of the Bavarian Center for Battery Technology (BayBatt) for the funding the ongoing research. 
\section{Author Contributions}
S.V.S.G collected data, prepared, trained and evaluated the ML and meta learning models. L.W. played a key role in data collection and curation. C.K. supervised and conceived this work.  
\section{Competing Interests}
The authors declare no competing financial interest.

\end{document}


\section{Supplementary Discussion}

While training MT-ML and meta learner models, we divided total dataset into two parts, namely train (80 percent) and validation (20 percent) respectively. We further subdivided the train into 5 parts. Each sub-part contains training and testing set. One separate model is prepared for each of them, using hyper-parameter tuning. Since our models are trained on multiple-anodes and material types, we compared our results for all of them. 

\begin{table}[hbtp]
\centering
\label{tab:specific_capacity}
\caption{Specific capacity (mAh/g) prediction across various charge carriers (positive electrode materials) and organic material classes (molecule/polymer) which is evaluated using metrics (MAE, R\textsuperscript{2}, and RMSE). These metrics are derived from cross-validation results (averaged across all test-sets of MT-ML models) and meta-learner models.}
\resizebox{1\textwidth}{!}{
\begin{tabular}{lrrrrrrrr}
\toprule
{Charge carrier (+ve)} & \multicolumn{4}{c}{Cross validation} & \multicolumn{4}{c}{Meta} \\
{} & {DP} & {MAE} & {R2} & {RMSE} & {DP} & {MAE} & {R2} & {RMSE} \\
\midrule
\multicolumn{9}{c}{Polymer} \\ 
\midrule
Al & 3 & 83.21 & -17.63 & 88.69 & 3 & 20.16 & -0.38 & 24.11 \\
K & 1 & 22.37 & & 22.37 & 1 & 1.22 & & 1.22 \\
Li & 117 & 60.77 & 0.55 & 102.43 & 117 & 26.22 & 0.92 & 44.27 \\
Mg & 5 & 32.04 & 0.66 & 44.61 & 5 & 5.56 & 0.99 & 8.69 \\
Na & 33 & 14.77 & 0.91 & 32.41 & 33 & 8.83 & 0.99 & 10.58 \\
Zn & 3 & 56.31 & 0.65 & 63.72 & 3 & 30.20 & 0.85 & 42.35 \\
\midrule
\multicolumn{9}{c}{Molecule} \\ 
\midrule
K & 2 & 25.91 & 0.36 & 33.13 & 2 & 8.52 & 0.96 & 8.52 \\
Li & 153 & 35.31 & 0.81 & 55.96 & 153 & 14.82 & 0.97 & 21.58 \\
Mg & 2 & 39.69 & -0.15 & 49.84 & 2 & 24.26 & 0.65 & 27.50 \\
Na & 51 & 23.40 & 0.88 & 37.11 & 51 & 13.14 & 0.98 & 16.97 \\
\bottomrule
\end{tabular}}
\end{table}

\begin{table}[hbtp]
\centering
\label{tab:voltage}
\caption{Voltage (V) prediction across various charge carriers (positive electrode materials) and organic material classes (molecule/polymer) which is evaluated using metrics (MAE, R\textsuperscript{2}, and RMSE). These metrics are derived from cross-validation results (averaged across all test-sets of MT-ML models) and meta-learner models.}
\resizebox{1\textwidth}{!}{
\begin{tabular}{lrrrrrrrr}
\toprule
{Charge carrier (+ve)} & \multicolumn{4}{c}{Cross validation} & \multicolumn{4}{c}{Meta} \\
{} & {DP} & {MAE} & {R2} & {RMSE} & {DP} & {MAE} & {R2} & {RMSE} \\
\midrule
\multicolumn{9}{c}{Polymer} \\ 
\midrule
Al & 2 & 0.84 & -307.45 & 0.88 & 2 & 0.36 & -55.28 & 0.37 \\
K & 1 & 0.20 & & 0.20 & 1 & 0.11 & & 0.10 \\
Li & 86 & 0.31 & 0.88 & 0.49 & 86 & 0.07 & 1.00 & 0.10 \\
Mg & 4 & 0.25 & -14.01 & 0.35 & 4 & 0.07 & 0.15 & 0.10 \\
Na & 31 & 0.08 & 0.97 & 0.14 & 31 & 0.07 & 0.99 & 0.10 \\
Zn & 2 & 0.67 & -52.10 & 0.73 & 2 & 0.17 & -3.92 & 0.22 \\
\midrule
\multicolumn{9}{c}{Molecule} \\ 
\midrule
Al & 1 & 0.38 & & 0.37 & 1 & 0.04 & & 0.00 \\
K & 2 & 0.37 & 0.36 & 0.42 & 2 & 0.04 & 0.99 & 0.00 \\
Li & 80 & 0.24 & 0.89 & 0.42 & 80 & 0.08 & 0.99 & 0.14 \\
Mg & 1 & 0.73 & & 0.73 & 1 & 0.22 & & 0.22 \\
Na & 33 & 0.19 & 0.78 & 0.35 & 33 & 0.06 & 0.99 & 0.10 \\
Zn & 3 & 0.86 & -43.57 & 0.98 & 3 & 0.13 & -0.15 & 0.14 \\
\bottomrule
\end{tabular}}
\end{table}

\begin{figure}[hbt]
 \includegraphics{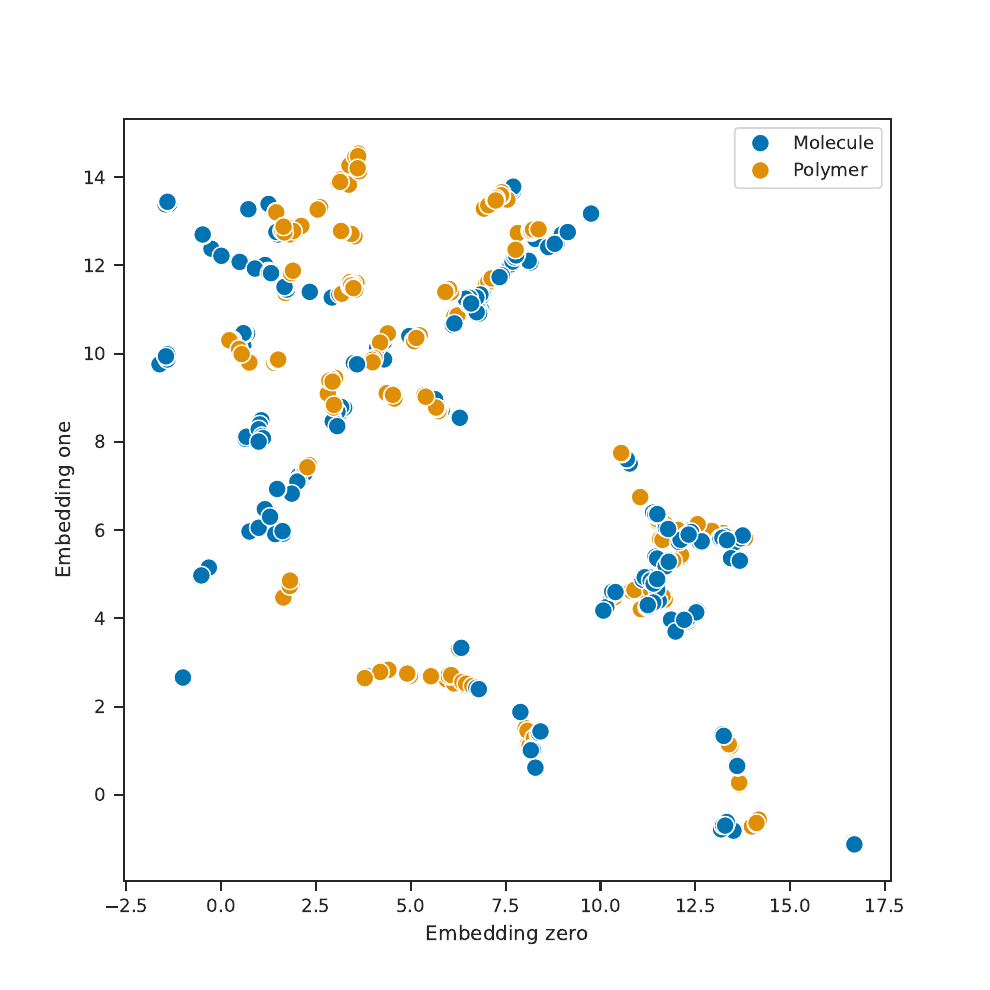}
  \caption{UMAP plots for our complete dataset plotted with respect to organic material classes as shown in the legend.}
  \label{fig:umap_plots}
\end{figure}

\begin{figure}[hbt]
 \includegraphics{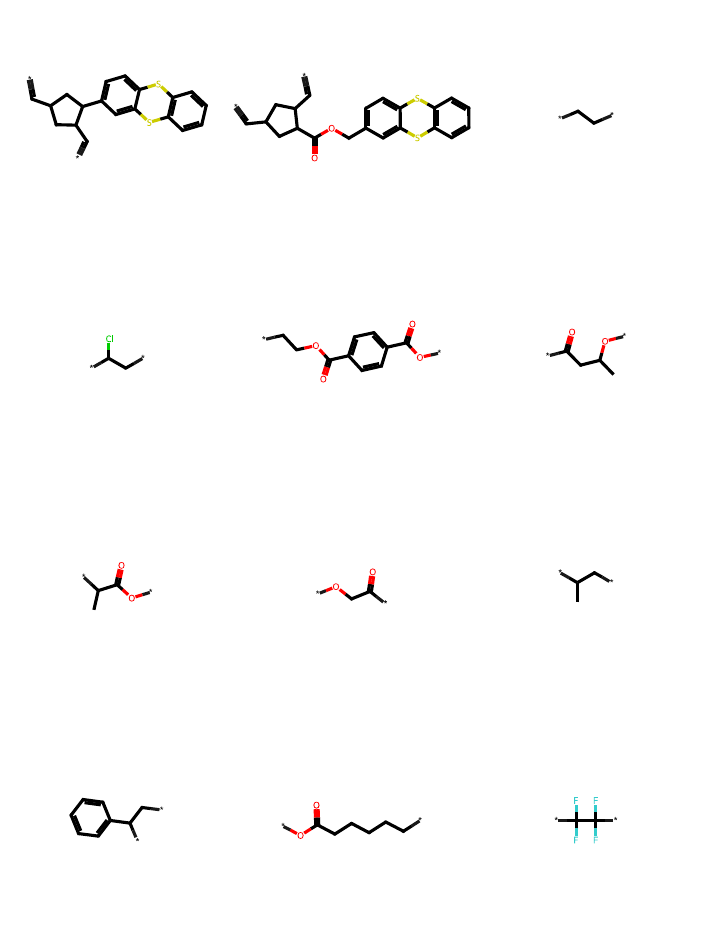}
  \caption{Reference organic polymers that are used to enumeratively add or replace moieties.}
  \label{fig:reforg}
\end{figure}

\begin{figure}[hbt]
 \includegraphics{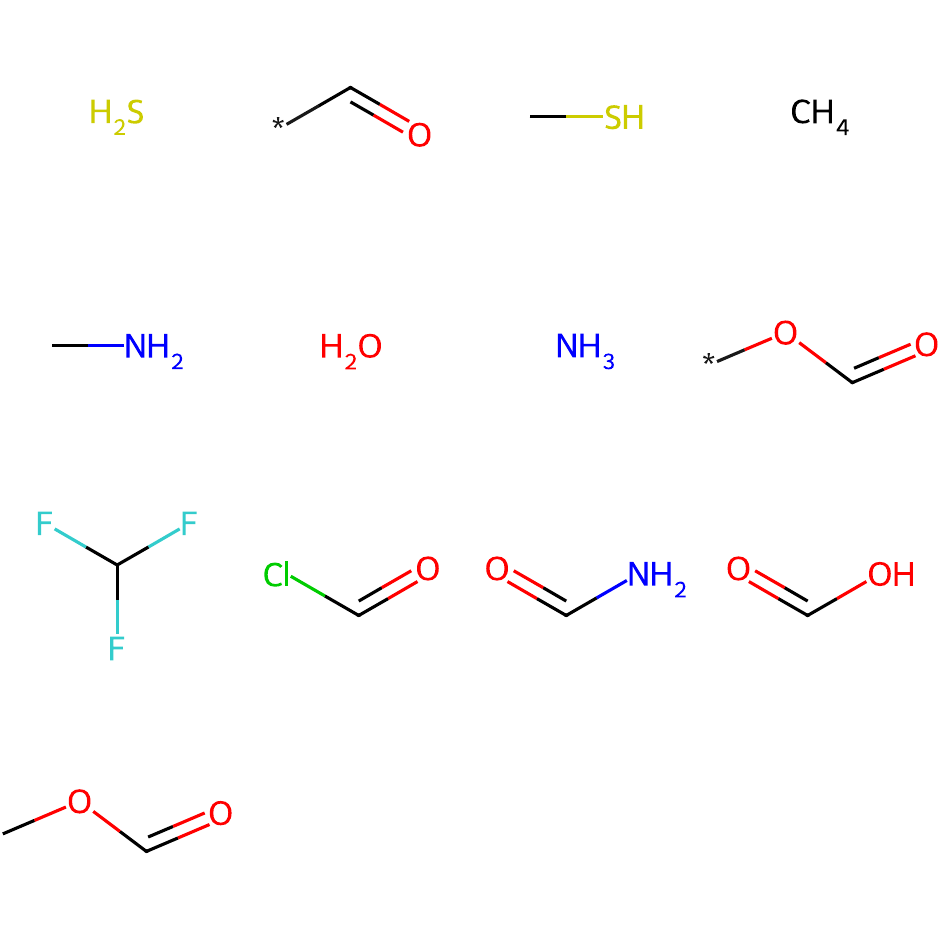}
  \caption{Moieties used to add or replace chemical bonds in reference organic materials.}
  \label{fig:moeties}
\end{figure}

\begin{figure}[hbt]
 \includegraphics{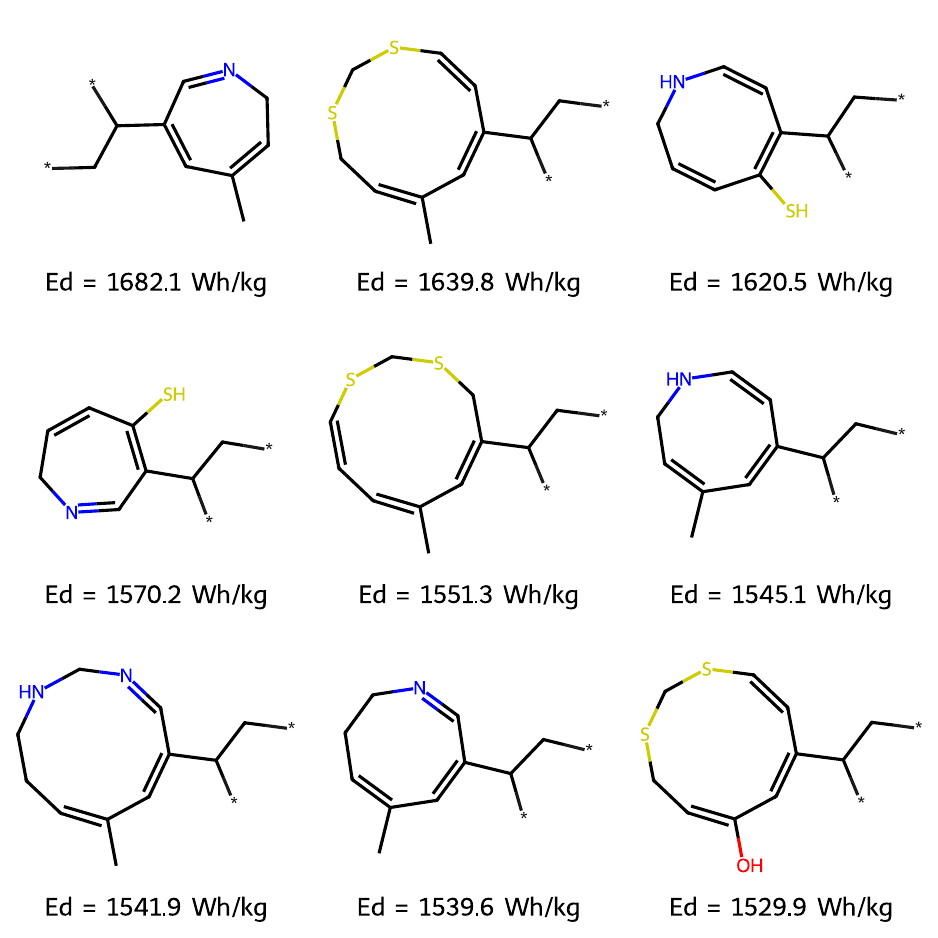}
  \caption{Top 9 candidates with the highest energy density curated from the 3 libraries totaling 1.8 million candidates.}
  \label{fig:ed_best_candidates}
\end{figure}

\begin{figure}[hbt]
 \includegraphics{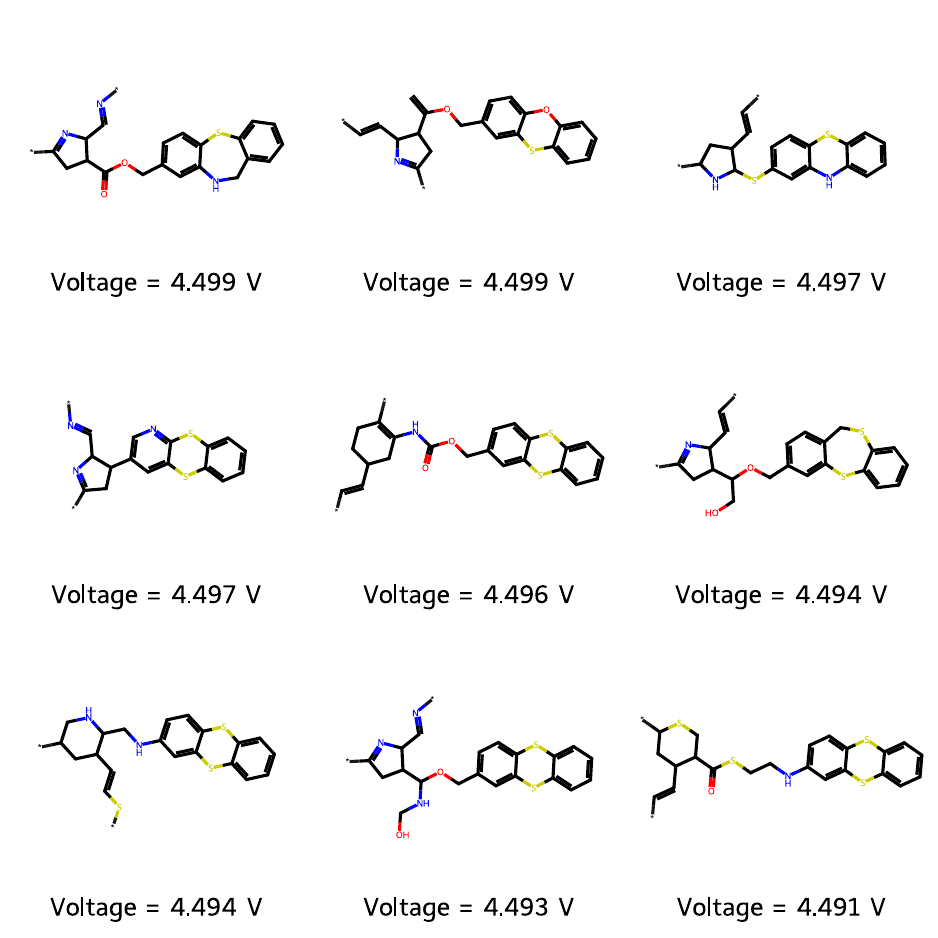}
  \caption{Top 9 candidates with the highest voltage curated from the 3 libraries totaling 1.8 million candidates.}
  \label{fig:voltage_best_candidates}
\end{figure}

\begin{figure}[hbt]
 \includegraphics{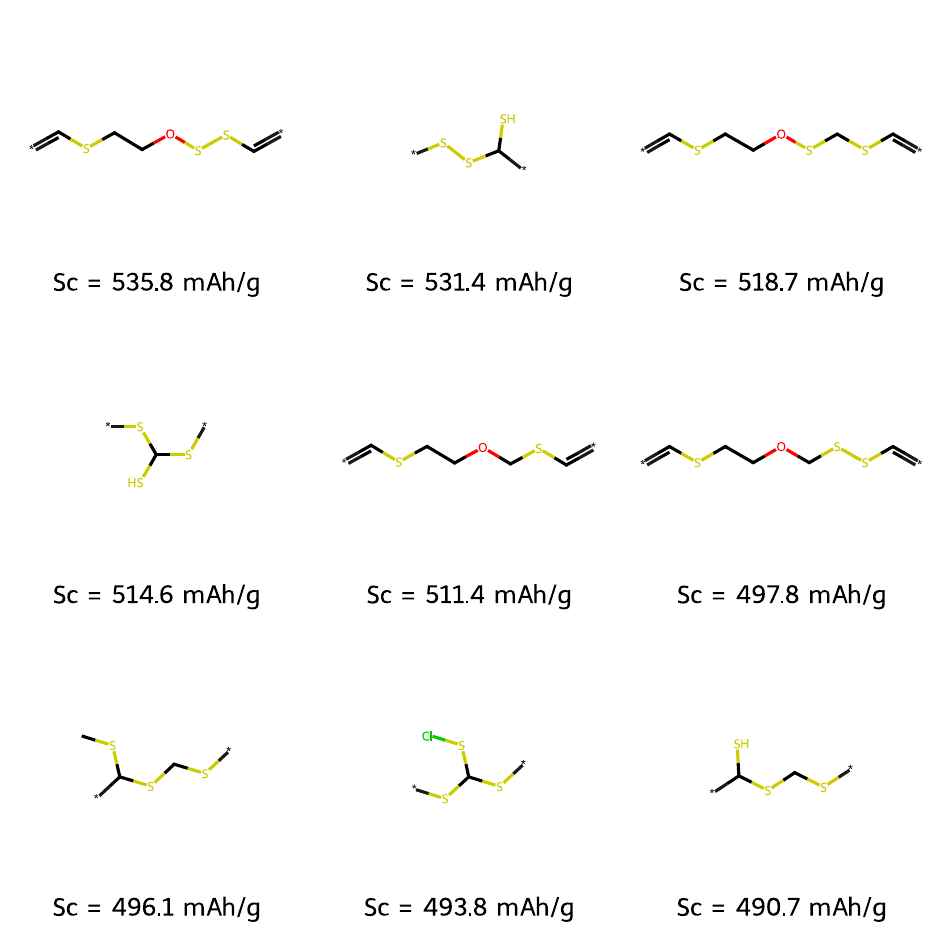}
  \caption{Top 9 candidates with the highest specific capacity curated from the 3 libraries totaling 1.8 million candidates.}
  \label{fig:sc_best_candidates}
\end{figure}